\RequirePackage{fix-cm}

\documentclass[twocolumn]{svjour3}
\smartqed  
\usepackage{epsfig}  
\usepackage{graphicx}
\usepackage{latexsym}
\usepackage{mathrsfs}
\usepackage{amssymb,amsbsy}
\usepackage{amsmath}
\usepackage{amsfonts}
\usepackage[authoryear]{natbib}
\RequirePackage{fix-cm}
\journalname{}
\newcommand{\D}{\Delta}
\renewcommand{\l}{\lambda}
\newcommand{\apj}{\it Astrophys.~J.}

\newcommand{\apss}{\it Astrophys. Space Sci.}

\newcommand{\nature}{\it Nature}

\newcommand{\araa}{\it Ann. Rev. Astron. Astrophys.}

\def\be{\begin{equation}}
\def\ee{\end{equation}}
\def\fr{\frac}

\begin{document}

\title{Prospect for UV observations from the Moon. II. Instrumental Design of an Ultraviolet Imager LUCI}

\author{Joice Mathew\and Ajin Prakash \and Mayuresh Sarpotdar   \and A.G. Sreejith \and Nirmal K. \and S. Ambily \and  Margarita Safonova \and Jayant Murthy\and Noah Brosch}

\institute{Joice Mathew\and Ajin Prakash \and Mayuresh Sarpotdar   \and A.G. Sreejith \and Nirmal K. \and S. Ambily \and  Margarita Safonova \and Jayant Murthy\at
Indian Institute of Astrophysics, Koramangala 2nd block, Bangalore, 560034, India \\
              \email{joice@iiap.res.in}        \\
           \and
Noah Brosch\at
The Wise Observatory and the Dept. Of Physics and Astronomy, Tel Aviv University, Tel Aviv 69978, Israel}

\date{Received: date / Accepted: date}

\maketitle

\begin{abstract}
We present a design for a near-ultraviolet (NUV) imaging instrument which may be flown on a range of available platforms, including high-altitude balloons, nanosatellites, or space missions. Although all current UV space missions adopt a Ritchey-Chretain telescope design, this requires aspheric optics, making the optical system complex, expensive and challenging for manufacturing and alignment. An all-spherical configuration is a cost-effective and simple solution. We have aimed for a small payload which may be launched by different platforms and we have designed a compact, light-weight payload which will withstand all launch loads. No other UV payloads have been previously reported with an all-spherical optical design for imaging in the NUV domain and a weight below 2 kg. Our main science goal is focussed on bright UV sources not accessible by the more sensitive large space UV missions. 

Here we discuss various aspects of design and development of the complete instrument, the structural and finite-element analysis of the system performed to ensure that the payload withstands launch-load stresses and vibrations. We expect to fly this telescope -- Lunar Ultraviolet Cosmic Imager (LUCI) -- on a spacecraft to the Moon as part of the Indian entry into Google X-Prize competition. Observations from the Moon provide a unique opportunity to observe the sky from a stable platform far above the Earth's atmosphere. However, we will explore other opportunities as well, and will fly this telescope on a high-altitude balloon later this year.
\end{abstract}


\section{Introduction}

Despite the fact that interesting and unique science may come through observations in the UV domain, this area remains largely unexplored \citep{Shustov}. Though large space observatories provide strong and important science outcome, they are also extremely expensive, and can only be afforded by governmental agencies. However, significant science can still be achieved by small telescopes which can be realized in a cost-effective way by launching them on a range of easily accessible near-space platforms, such as sounding rockets, high-altitude balloons, CubeSats, nanosatellites, etc.  \citep[e.g.][]{Brosch,Broschsmall}. Note that the first UV spectrum of a quasar was obtained by a 16-inch telescope on a short rocket flight \citep{Davidsen}.

Most of the previous and existing UV space missions have severe brightness limits (NUV mag $< 10$) due to detector  safety constraints (e.g. UVIT and GALEX), and therefore, bright objects are usually excluded from observations. Our goal is to develop a UV payload ready to fly on a range of available platforms, with  the science objective to observe bright objects in the near-UV (NUV: 200 -- 300 nm), such as hot massive stars, bright transients, solar system bodies, etc. \citep{Broschsmall,Safonova}. The main design considerations for such an instrument are the capability to observe bright sources in the NUV, light weight and compactness, cost-effective development, and space qualification.

The current trend in development of compact wide-field telescopes is the Ritchey-Chretien (RC) design \citep[e.g.][]{Roming}. However, the hyperbolic mirrors in such design increase the cost and complexity of the system. We have selected an all-spherical catadioptric design, which reduces the manufacturing complexity and cost. The optical system is a two-mirror configuration with spherical mirrors and a double-pass corrector lens. We discuss in detail the optical and mechanical design of the instrument in the following sections. The structural design is followed by a finite element analysis (FEA) performed to ensure that the payload withstands all launch loads and vibrations.

Currently, we collaborate with the Team Indus, an Indian contestant for the Google Lunar X~PRIZE\footnote{The Google Lunar XPRIZE is the international competition with \$30 million incentive sponsored by Google and operated by the XPRIZE Foundation; official website {\tt http://www.googlelunarxprize.org/}.}, competition to send this UV telescope -- Lunar Ultraviolet Cosmic Imager (LUCI) -- to the Moon. In Safonova et al. (2014), hereafter called paper~I, we have described the proposal, science goals and the original design of LUCI. Unfortunately, the extremely severe mission constraints on weight and volume of payloads reduced our allowances, and we had to dramatically redesign the instrument in order to be able to keep up with the proposed science goals and mission requirements. We selected the option of a NUV telescope with spherical optical elements and a UV-enhanced CCD detector, which turned out to be a cost-effective, simple and quick solution to realize in the potential UV payloads. The reduction of the weight to under 2 kg, and the volume to under 0.01 m$^{3}$, made LUCI a unique space instrument as no other NUV telescope have been reported with an all-spherical optical design for imaging with such compact dimensions. LUCI will be mounted on the lunar lander as a transit telescope, and will perform a survey of the available sky from the surface of the Moon, with the aim to detect bright UV transients. Observations from the Moon provide a unique opportunity to observe the UV sky from a stable platform far above the Earth's atmosphere.

\section{Instrument Overview}

\begin{figure}[ht!]
\begin{center}
\includegraphics[scale=0.55]{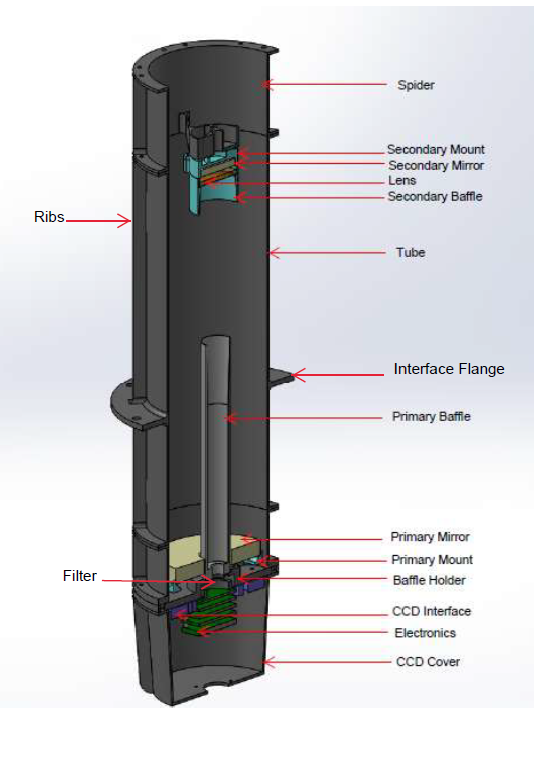}
\end{center}
\caption{LUCI Cross-Section.}
\label{fig:LUCI cross-section}
\end{figure}

The instrument is a UV telescope, where all optical elements are spherical. It will be used to study the variability and environment of bright UV sources by acquiring photometric time-series in the 200 -- 300 nm wavelength range. It will acquire images at a fast frame rate, and analyze each frame looking for brightness variations across the frames. The transient events will be stored on-board, and send back to Earth whenever the radio link is available. The cross-section of the instrument is shown in Fig.~\ref{fig:LUCI cross-section}, and technical specifications are given in Table~\ref{table:instrument details}.

\begin{table}[h]
\caption{Instrument details}
\begin{center}
\begin{tabular}{ll}
\hline 
\rule[-1ex]{0pt}{3.5ex} Instrument & UV Imager (LUCI) \\
\rule[-1ex]{0pt}{3.5ex} Telescope type & Spherical catadioptric  \\
\rule[-1ex]{0pt}{3.5ex} Dimension (L $\times$ D)& $450 \times 150$ mm \\
\rule[-1ex]{0pt}{3.5ex} Weight & $ 1.85 $ kg \\
\rule[-1ex]{0pt}{3.5ex} Power & $ < 5$  W \\
\rule[-1ex]{0pt}{3.5ex} Aperture diameter  &  80 mm  \\
\rule[-1ex]{0pt}{3.5ex} Focal length & 800.69 mm  \\
\rule[-1ex]{0pt}{3.5ex} Field of view & $27.6^{\prime}\times 20.4^{\prime}$ \\
\rule[-1ex]{0pt}{3.5ex} Detector & UV-sensitive CCD \\
\rule[-1ex]{0pt}{3.5ex} Sensor format (H$\times$V)& $1360 \times 1024$ \\
\rule[-1ex]{0pt}{3.5ex} Pixel size & $4.65\times 4.65\,\mu$m \\
\rule[-1ex]{0pt}{3.5ex} Pixel scale & $1.2^{\prime\prime}$/pixel \\
\rule[-1ex]{0pt}{3.5ex} Spatial resolution & $\sim 5^{\prime\prime} $\\
\rule[-1ex]{0pt}{3.5ex} Band of operation & $200-320$ nm\\
\rule[-1ex]{0pt}{3.5ex} Limiting magnitude & 12 AB\\
\rule[-1ex]{0pt}{3.5ex} Bright limit & 2 AB\\
\rule[-1ex]{0pt}{3.5ex} Minimum exposure time & 0.08 sec \\
\hline
\end{tabular}
\label{table:instrument details}
\end{center}
\end{table}

The mirrors are coated with an Al reflective surface on Zerodur substrate, and a protective MgF$_2$ coating to preserve the reflectivity in the NUV, which gives a broad bandpass of 120 -- 900 nm. A double-pass lens is used to correct for aberrations. The lens is made from fused silica, which has more than 90\% transmission down to 195 nm\footnote{https://www.newport.com/n/optical-materials}. In addition, both sides of the lens are coated with the anti-reflection MgF$_2$ coating to supress the reflection and thus maximize the transmission (up to 95\%) in the NUV band. The detector is a UV-sensitive CCD (Charge Coupled Device) with the response between $200 - 900$ nm; therefore we have placed a solar blind filter before the CCD (at a distance of 15 mm before the focal plane) to restrict the bandpass to $200 - 320$ nm. This UV filter has a peak transmission of $\geq 30\%$ at 250 nm and has an excellent cut-off in the visible and IR bands. 

An FPGA-based electronics processing board is developed in-house \citep{Sapotdar} to acquire the data from the detector, and for the subsequent on-board processing, compression and storage. We have designed the instrument by considering the strict weight and volume constraints -- the major issues in the inexpensive space platforms, and the possible launch-load vibrations. To meet the above requirements, we have implemented a tubular structure with ribs. To maintain the working temperature, heat strips will be provided along the ribs, and the instrument will be covered with Multi-Layered Insulation (MLI).

The additional strict constrain imposed by the lunar mission was the necessity to reduce the weight to below~2~kg. In LUCI implementation, the power and telemetry for the instrument will be provided by the lunar lander. The telescope will be mounted on the lunar lander in a storage bay, where temperature will be maintained between $24^{\circ}$C and $26^{\circ}$C. A one-time opening door is under design for the telescope to avoid any possible contamination from outside. 

\section{Optical design}

\begin{figure*}[ht]
\begin{center}
\includegraphics[scale=0.35]{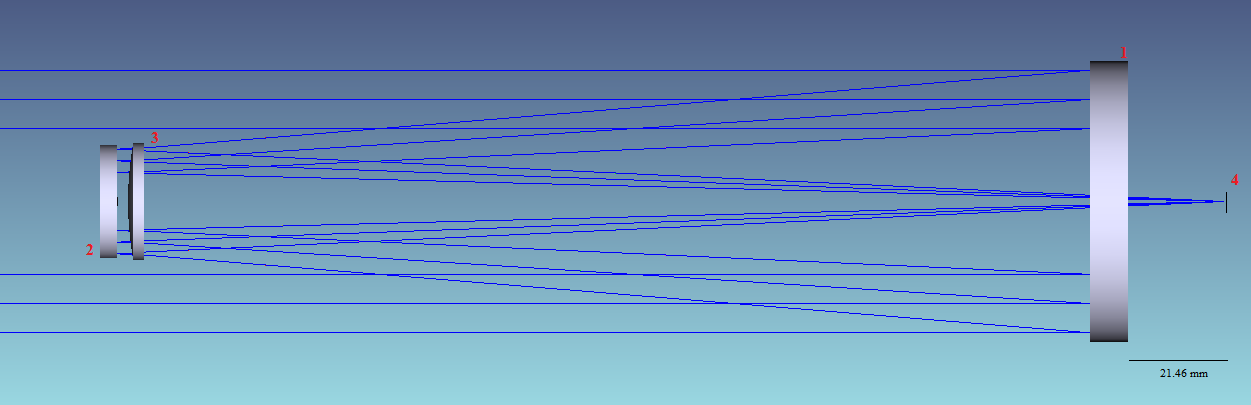}
\end{center}
\caption{Optical layout: 1. primary mirror; 2. secondary mirror; 3. meniscus lens; 4. focal plane.}
\label{fig:Optical Layout}
\end{figure*}

Current UV space missions use  Ritchey-Chr\'{e}tien (RC) based telescope design \citep{Cao,Kumar,Jeong}. The RC based telescope configuration consists of two hyperbolic mirrors; manufacturing and testing of these aspheric surfaces are relatively complex and expensive. Our main design goal was to come up with a cost-effective alternative, and it is achieved by a design consisting of the spherical primary and secondary mirrors, and a spherical meniscus double-pass lens, placed in front of the secondary mirror to reduce the spherical aberrations. The advantages of telescope configurations employing only spherical surfaces are low cost and the ease of manufacturing and alignment. The optical design was carried out using Zemax, and the variables used for the design optimization  are the radius of curvature of the optical elements, and the spacing between the elements. We present the optical layout of the telescope in Fig.~\ref{fig:Optical Layout}. 
Parallel rays from the source, after reflecting from the primary mirror, pass through the double-pass lens and fall on the secondary mirror. After reflecting from the secondary mirror, the rays pass through the lens once again and get focused at the focal plane. The center spacing between the primary mirror and lens is 272.53 mm, and between the lens and secondary mirror 3.00 mm. The meniscus lens with a semi-diameter of 16.77 mm has the central thickness of 3.00 mm. The focal plane is set at a distance of 310 mm from the lens. The paraxial image height is 3.24 mm. The ray tracing parameters for the optical system are given in Table~\ref{table:optical parameters}.

\begin{table*}[ht]
\begin{center}
\caption{Optical parameters}
\begin{tabular}{llllc}
\hline
Surface & Radius (mm) & Thickness (mm) & Optical material & Semi Diameter (mm)\\
 \hline
\rule[-1ex]{0pt}{3.5ex}  Primary Mirror & $-914.30$ & $-272.53$ & Zerodur Mirror & 40.00\\
\rule[-1ex]{0pt}{3.5ex}  Lens & 90.09 & $-3.00$ & Fused Silica & 16.77\\
  & 91.31 & $-3.00$ &  & 16.46\\
\rule[-1ex]{0pt}{3.5ex}   Secondary Mirror & $-865.53$ & 3.00 & Zerodur Mirror & 16.27\\
\rule[-1ex]{0pt}{3.5ex}   Lens & 91.31 & 3.00 & Fused silica  & 16.16\\
  & 90.09 & 310.00&  & 16.25\\
\rule[-1ex]{0pt}{3.5ex}   Image & Infinity &  &   & 3.24\\
  \hline
  \end{tabular}
\label{table:optical parameters}
\end{center}
\end{table*}

\begin{figure}[h!]
\begin{center}
\includegraphics[scale=0.57]{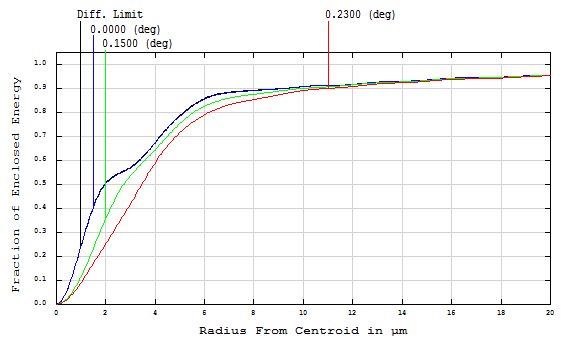}
\caption{Diffraction-encircled energy for on-axis and off-axis fields. Colours indicate different fields, with offset from the optical axis shown in the plot.}
\label{fig:Encircled Energy}
\end{center}
\end{figure}

\begin{figure}[h!]
\begin{center}
\includegraphics[scale=0.59]{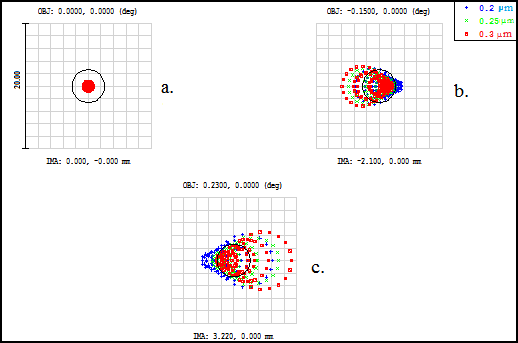}
\end{center}
\caption{Spot diagram for on-axis (a.) and off-axis (b.and c.) fields. Colours indicate the wavelength: blue for 0.2 $\mu m$, green for 0.25 $\mu m$, and red for 0.3 $\mu m$. Black circle  shows the Airy disk with radius 2.6 $\mu m$.}
\label{fig:spot}
\end{figure}

It is ensured that at least 80\% of encircled energy (Fig.~\ref{fig:Encircled Energy}) falls within an area of $4\times 4$ pixel for the highest degree off-axis field,  after considering the manufacturing and alignment tolerances. The spot diagram is shown in Fig.~\ref{fig:spot}. For the on-axis field, the  spot radius is within the Airy disk radius, and for the highest field, RMS spot radius is 3.49 $\mu m$.

\subsection{Scattering analysis}

\begin{figure}[h!]
\begin{center}
\includegraphics[scale=0.35]{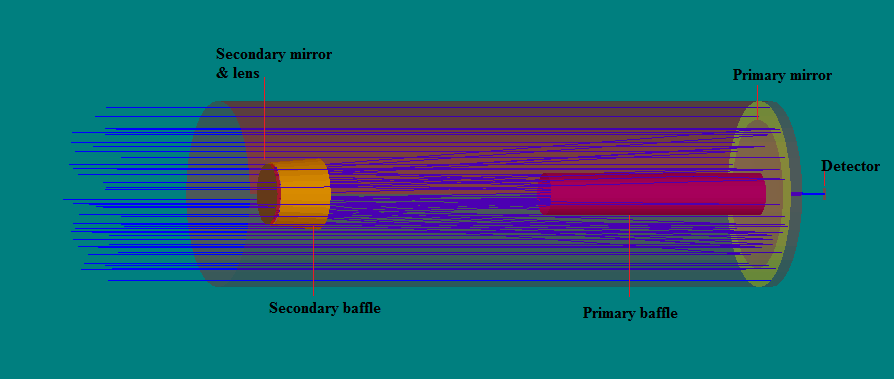}
\end{center}
\caption{The stray light analysis model of LUCI in Zemax.}
\label{fig:stray}
\end{figure}

Primary and secondary baffles are implemented in LUCI to reduce the effect of the stray light at the focal plane. We have developed an iterative algorithm to find the optimum sizes of the baffles based on the numerical technique mentioned in \citep{Moore}. 
Required structure of the baffles, sources and detectors, and the performance of designed baffles have been evaluated using the non-sequential ray tracing method in Zemax optical design software (Fig.~\ref{fig:stray}). Results show that designed baffles block the direct rays of (zeroth-order) stray light reaching the image plane, and control the higher-order stray light that originates due to the finite reflective nature of the primary mirror baffle structure.

A scattering model has been done in the Zemax non-sequential mode. The walls of the telescope tubes, baffles and mirror mounts are black painted with Aeroglaze Z306 (which has more than 95\% absorptivity) to suppress the scattered light. To perform the scattering analysis, we assumed all the telescope tubes and baffles to have a Lambertian approximation. We have used ABg \citep{Freniere} model to characterize the Bi-Directional Scatter Distribution Function (BSDF) of the mirrors with the $A,B,g$ values of $0.0015$, $0.001$ and $2$, respectively. These $A,B,g$ values give a Total Integrated Scatter (TIS) of around 0.015, equivalent to the mirrors with a surface roughness of around 25 \AA\, rms.

\begin{figure}[h!]
 \includegraphics[scale=0.25]{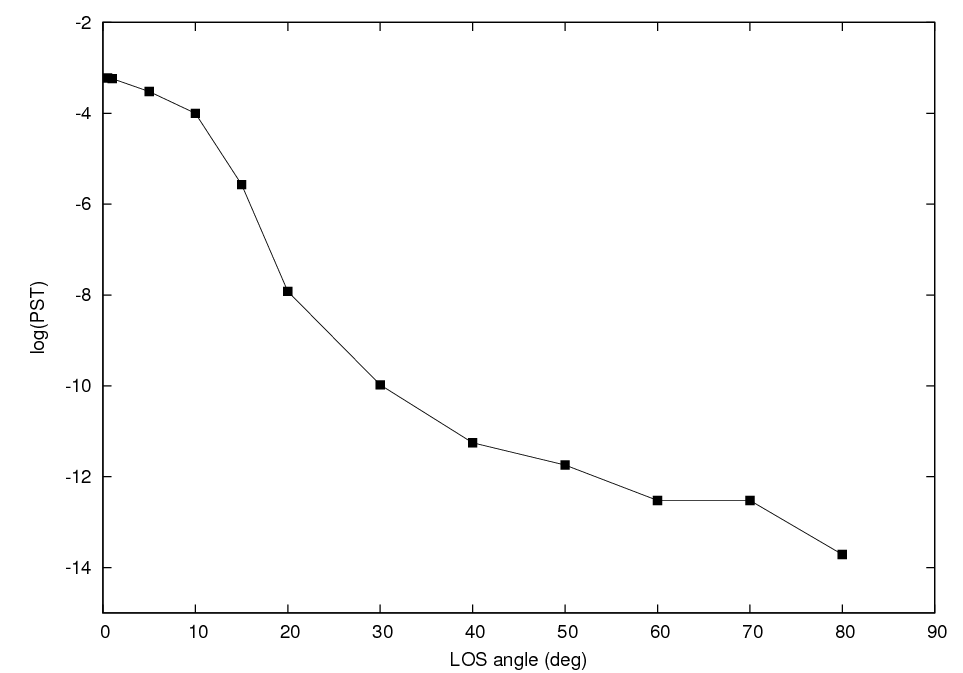}
 \caption{Modelled LUCI PST for the stray light from a near-field point-like source.}
\label{fig:PST}
\end{figure}

 A tilted collimated source was constructed using the source ellipse object with the diameter larger than the entry port, ensuring the complete filling of the entry port at maximum tilt angle. At the image plane, a rectangle surface detector having sides of $6.4\times 4.8$ mm is placed to quantify the total light power received at the image plane. One million rays from the incident collimated light having 1 watt of total power equally divided among them was non-sequentially traced through the telescope model. The criterion to terminate the tracing of a ray was kept at $10^{-8}$ times the ray source input power. We have calculated the total flux reaching the detector plane for different angles of incidences, and modelled the point-source transmission function (PST), which, for a given incidence angle, is the ratio between incidence irradiance of a point source at instrument entrance level, and transmitted irradiance at the focal plane level. The PST is shown in Fig.~\ref{fig:PST}, where LOS is the line-of-sight angle in degrees. 
 
Ghost image analysis was also performed in Zemax using ghost image focus generator option. We found that no ghost image is formed at the focal plane, and the nearest ghost image focus is 70 mm before the detector plane. Since the lens is AR-coated (with 95\% transmission in NUV and a reflectivity less than 0.5\%), the contribution of flux due to ghost rays is small, around 0.2 \% of the actual image flux on the detector plane.

\subsection{Tolerance analysis}

 To evaluate the performance of the opto-mechanical system, we carried out tolerance analysis. We have combined sensitivity analysis and Monte Carlo simulations using Zemax to investigate the degradation of the encircled energy. The tolerances  for manufacturing and alignment are shown in Table~\ref{table:tolerance values}. Back focal length of the telescope is used as the compensator. Under ideal conditions, 80\% of the encircled energy should fall within $3\times 3$ pixels. With our tolerance allocations, the encircled energy falls within $4\times 4$ pixels, which is still within the acceptable optical performance. Therefore, the designed system meets the required imaging criteria after considering the manufacturing and assembling precision limits.

\begin{table*}[ht]
\begin{center}
\caption{Tolerance allocation on manufacturing and alignment}
\begin{tabular}{lllc}
\hline
Tolerance term & sub tolerance term & objects & tolerances\\
\hline
\rule[-1ex]{0pt}{3.5ex} Manufacture & Radius of curvature & Primary Mirror & 1\% \\
   &  & Secondary Mirror & 1\% \\
  &  & Lens & 0.1\% \\
  & Thickness ($\mu$m) & Primary Mirror & $\pm 100$\\
  &  & Secondary Mirror &$\pm 100$ \\
  &  & Lens & $\pm 50$ \\
  & Decenter in X $\&$ Y ($\mu$m) & Primary Mirror & $\pm 50$\\
  &  & Secondary Mirror &$\pm 50$ \\
  &  & Lens & $\pm 50$ \\
  & Tilt in X $\&$ Y ($^{\prime\prime}$)  & Primary Mirror & 60\\
  &  & Secondary Mirror &60 \\
  &  & Lens & $60$ \\
  & Surface irregularity  & Primary Mirror & $\Lambda/6$\\
  &  & Secondary Mirror &$\Lambda/6$ \\
  &  & Lens & $\Lambda/6$ \\
\rule[-1ex]{0pt}{3.5ex} Alignment & Decenter in X $\&$ Y ($\mu$m) & Primary Mirror & $\pm 50$\\
            &  & Secondary Mirror &$\pm 50$ \\
            &  & Lens & $\pm 50$ \\
             & Tilt in X $\&$ Y ($^{\prime\prime}$)  & Primary Mirror & 60\\
             &  & Secondary Mirror &60 \\
             &  & Lens & $60$ \\
\hline
\end{tabular}
\label{table:tolerance values}
\end{center}
\end{table*}

\section{Mechanical Design}

\begin{figure*}[ht]
\begin{center}
\includegraphics[width=1\textwidth]{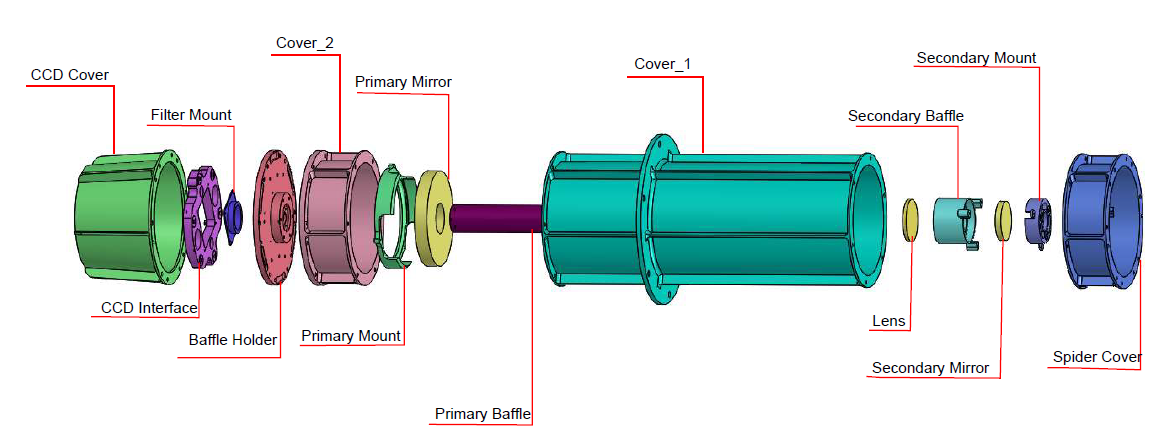}
\end{center}
\caption{Instrument structural layout.}
\label{fig:LUCI Structure}
\end{figure*}

The instrument is designed to meet such requirements as light weight, small volume, and structural stability to be able to  withstand all launch-load vibrations. The most stringent requirements for launch vehicle platforms \citep{Ariane} are the following: the natural frequency must be above 100 Hz, and the instrument should be able to withstand shock loads of up to 5 g. To achieve this, a high stiffness-to-mass ratio is adapted in the design by using six ribs along the telescope cover. Aluminum 6061-T6 is used for the tubular structure of the telescope, due to its high strength-to-density ratio, low outgassing properties, easy machinability and low cost. Invar36 is used for the mirrors and lens mounts due to its very low thermal expansion properties. The properties of the different materials that we have used are shown in Table~\ref{table:Material Properties}. We used the mechanical design software SolidWorks for the CAD (Computer Aided Design) model of the instrument. The exploded view of the instrument is shown in Fig.~\ref{fig:LUCI Structure}.

\begin{table}[h!]
\begin{center}
\caption{Properties of the materials used}
\begin{tabular}{llll}
\hline
Material Property & AA-6061&Invar36&Ti-6Al-4V\\
 \hline
Young's Modulus (MPa) &  68900 &148000&115000 \\
Density (tonne/mm$^3$) & 2.7$\times 10^9$ &8.05$\times 10^9$&4.30$\times 10^9$\\
Poisson's Ratio&  0.33 &0.29& 0.3\\
Tensile Yield Strength&320& 240& 880\\
\hline
\end{tabular}
\label{table:Material Properties}
\end{center}
\end{table}

\subsection{Structural Finite Element Analysis}

Finite element analysis (FEA) of the structure has to be performed in order to ensure that the payload can withstand all launch loads and vibrations \citep {LiE,JHLee}. Hypermesh
is used as the pre-processor for finite element modeling of the structure. The FEA model consists of 68,909 nodes and 66,038 elements for static and response analysis. Modal analysis was performed to determine the natural frequency of the structure. The static and buckling analysis was conducted with 25 g inertial fields in the $X$, $Y$ and $Z$ directions each. A response analysis was carried out for a frequency range of 5 Hz to 100 Hz.

\subsubsection{Modal Analysis}
 
\begin{figure}[ht!]
\begin{center}
\includegraphics[scale=0.55]{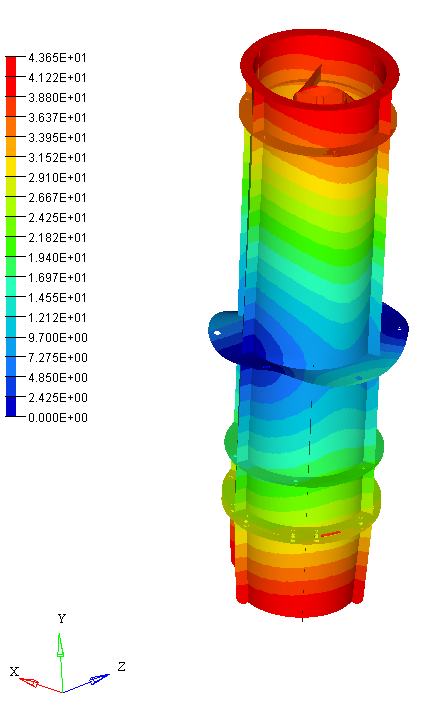}
\end{center}
\caption{Eigenfrequency of the first order resonance mode with natural frequency 202.25 Hz.}
\label{fig:First Mode}
\end{figure}

To avoid the resonance caused by the dynamic coupling between instrument and launch vibrations, the natural frequency of the instrument must be higher than that of a launch vehicle. Modal analysis was conducted to find the natural frequencies of the system. Because we want the instrument to be ready to fly on any platform, we made the design to have a natural frequency at least 1.5 times higher than required by most of the launch vehicles -- 100 Hz. The first six natural frequencies are given in Table~\ref{table:mode frequencies}, and the first mode of the telescope structure is shown in Fig.~\ref{fig:First Mode}.

\begin{table}[h]
\begin{center}
\caption{Modal frequencies}
\begin{tabular}{lc}
\hline
Modes & Frequencies \\
 \hline
First &  202.25 \\
Second &  202.35 \\
Third &  272.23 \\
Fourth &  276.47 \\
Fifth &  512.87 \\
Sixth &  513.45 \\
\hline
\end{tabular}
\label{table:mode frequencies}
\end{center}
\end{table}

\subsubsection{Static Analysis}

\begin{figure}[h!]
\begin{center}
\includegraphics[scale=0.4]{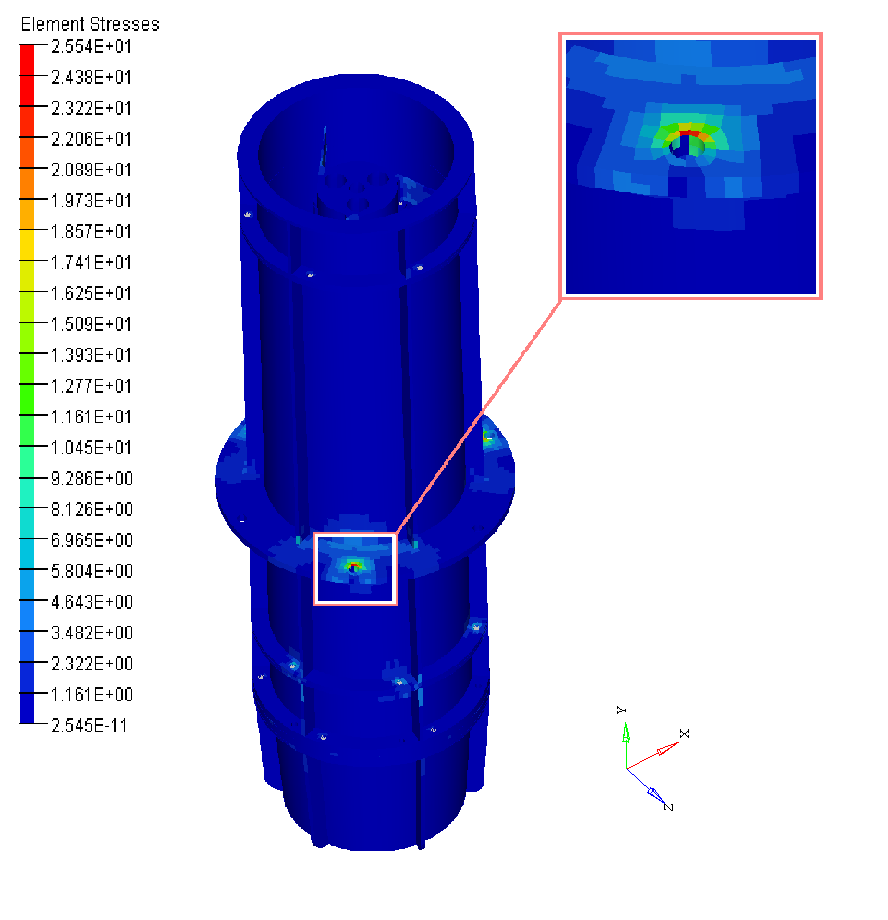}
\end{center}
\caption{25 g static analysis: minimum safety factor 10.76. Inset shows one of the three interface holes where maximum stress occurs.}
\label{fig:Static analysis}
\end{figure}

Static analysis is performed to assess the behavior and resistance of the structure to the quasi-static load. The static load would  induce stresses on the components due to the inertial effects. These stresses must be less than the yield strength of the material to avoid permanent deformation or damage to the components. Permanent deformation would result in bad optical quality and reduced performance, or sometimes in the complete failure of the system. Quasi-static load of 25 g was considered both in lateral $X/Z$ and longitudinal (axial) $Y$ directions for analysis. The maximum stress of 25.54 MPa during static loading of 25 g in the longitudinal direction is on the payload interface flange (Fig.~\ref{fig:Static analysis}), which is considerably smaller than the yield strength of Al 6061 (275 MPa). Therefore, the factor of safety (FOS) is  $FOS=275/25.54=10.76$ at the interface flange. Thus, we consider the structure to be safe for the quasi-static load during the launch.

\subsubsection{Frequency Response Analysis}

Frequency response analysis was conducted to find the response of the structure to the vibrations induced by the launch vehicle during the launch. The design must be able to sustain these vibrations without causing any permanent damage to the optical and electronic instruments on the payload. Sinusoidal frequency response analysis was carried out to determine the response of the structure for sinusoidal base excitation. The three bolt holes at the interface flange were excited from 5 Hz to 100 Hz in steps. Since the first natural frequency is above 200 Hz, there was no structural response in the given frequency range, i.e. up to 100 Hz.

\subsubsection{Buckling Analysis}

Buckling is a mode of failure of a mechanical component when it can no longer sustain applied compression forces, and hence, collapses or deforms. Due to the tubular shape of the instrument, it is susceptible to buckling. A buckling analysis was conducted with 25 g load applied both laterally and longitudinally. The bucking eigenvalues in three directions are shown in Table~\ref{table:Buckling_Eigenvalues}. Since the buckling eigenvalues are greater than one, we consider the structure to be safe for buckling for loads up to 25 g.

\begin{table}[h!]
\begin{center}
\caption{Buckling Eigenvalues}
\begin{tabular}{lc}
\hline
Direction & Eigenvalue \\
 \hline
X &  82.50 \\
Y &  149.36 \\
Z &  55.28 \\
\hline
\end{tabular}
\label{table:Buckling_Eigenvalues}
\end{center}
\end{table}       
\section{Detector and Electronics}

We are using a broadband ICX407BLA CCD from Sony, which is specially enhanced for its UV response and has quantum efficiency of more than 30\% in 200 -- 300 nm (Fig.~\ref{fig:filter}). The CCD is a diagonal 8-mm (type 1/2-inch) interline solid-state image sensor with 1360(H)$\times$1024(V) pixel format, and the pixel size is 4.65 $\mu m$. The detector electronics includes a generic (application-wise) in-house developed field-programmable gate array (FPGA) board \citep{Sapotdar} used as image processor board to generate the clocks and read the data, and a real-time processor system for image processing tasks of different levels. We have selected Spartan-6Q from Xilinx\footnote{http://www.xilinx.com} as our FPGA because of its processing capability and wide operating temperature range. We also implement a soft core microprocessor microblaze (Xilinx)  inside the FPGA. 
 
\begin{figure}[h]
\begin{center}
\includegraphics[scale=1.]{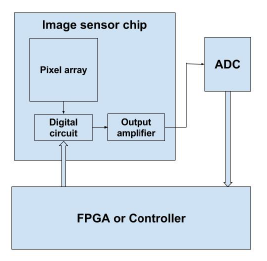} 
\end{center}
\caption{Electronics readout block diagram.}
\label{fig:Electronics readout block diagram}
\end{figure}

The sensor electronics readout and interface block diagram is shown in  Fig.~\ref{fig:Electronics readout block diagram}. There is a timing generator chip VSP01M01 along with the image sensor on the same  printed circuit board (PCB). This chip includes the digital circuitry to access the analog voltage from each pixel, and an analog to digital converter (ADC). It also generates image synchronization driver signals (frame-valid, line-valid, and pixel clock) which are received by the FPGA board. The required timing for various clock signals are to be programmed when circuit is switched on, and is done by microblaze in the FPGA. The FPGA board records the values of each pixel and  decodes the position of the pixel using the synchronization signals from the timing generator. The acquired frames will be scanned for brightness variations, and the transient events will be stored on-board to transmit the data via downlink. 

\section{Calibration}

\begin{figure}[h!]
\hspace{-0.2in}
\includegraphics[scale=0.25]{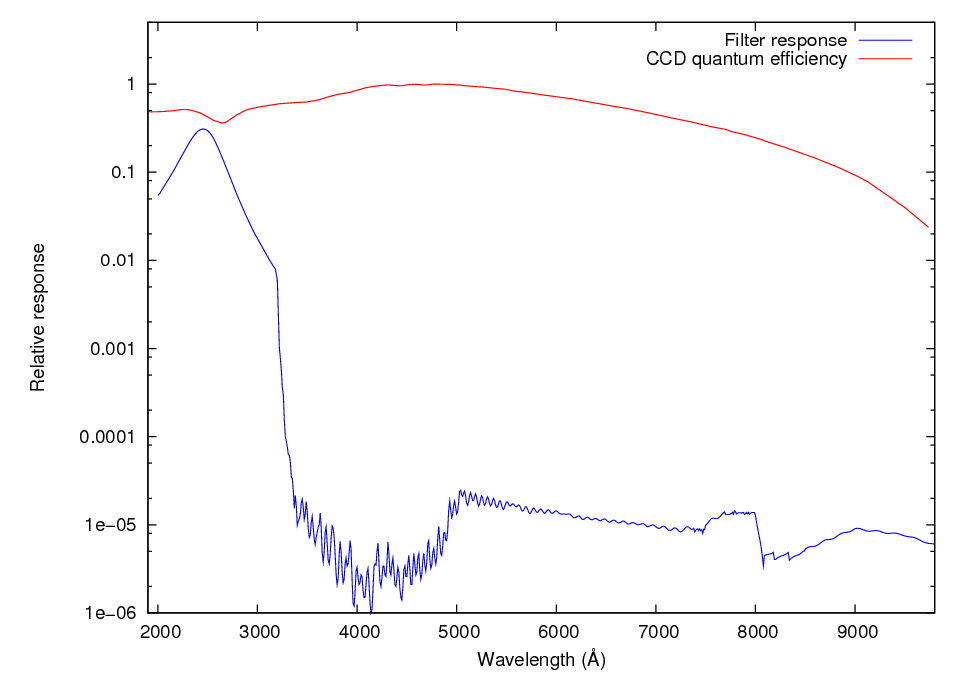} 
\caption{CCD quantum efficiency (Top curve) and filter transmission (Bottom curve) curves.}
\label{fig:filter}
\end{figure}

\begin{figure}[h!]
\hspace{-0.2in}
\includegraphics[scale=0.45]{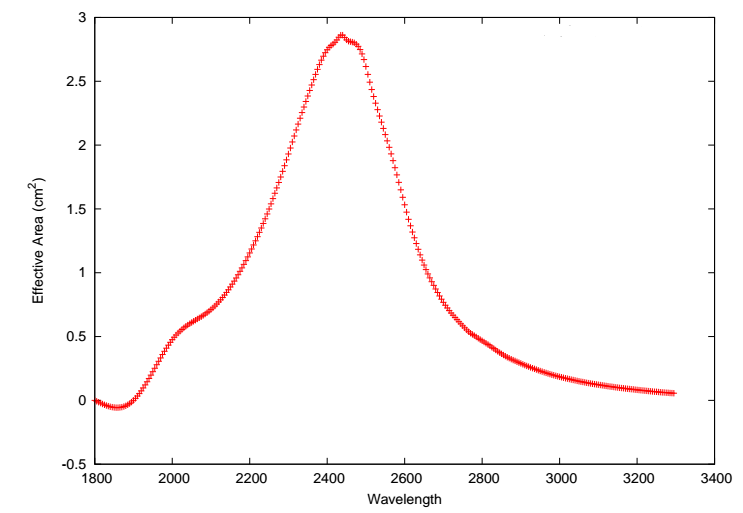} 
\caption{LUCI effective area in the NUV.}
\label{fig:effective}
\end{figure}

For the definition of LUCI photometric system, we adopt the AB magnitude system widely used in UV astronomy \citep{OkeGunn},
\be
m_{\rm LUCI}=-2.5\log{\langle F_{\nu}\rangle} -48.6\,,
\ee
where $\langle F_{\nu}\rangle$ is averaged across the bandpass monochromatic flux (ergs/sec/cm$^2$/Hz) at central (mean) wavelength $\lambda_0$,
\be
\lambda_0= \fr{\int \lambda R_{\lambda} d\lambda}{\int R_{\lambda} d\lambda}\,,\quad \langle F_{\nu}\rangle = \fr{\int f_{\nu} R_{\nu} 
d\nu}{\int R_{\nu} d\nu} \,, 
\ee
and where
\be
\langle F_{\lambda}\rangle = \fr{\int f_{\lambda} R_{\lambda} d\lambda}{\int R_{\lambda} d\lambda}\,\,,f_{\lambda}=f_{\nu}\fr{c}{\lambda^2_{\rm p} \cdot 10^8}\,\,,R_{\nu}=R_{\lambda}\,.
\label{eq:2}
\ee
Here $R_{\lambda}$ is the total system's response, $f_{\lambda}$ is the SED of a source, and $\lambda_{\rm p}$ is the pivot wavelength. We express the total system response, a constant characterizing the telescope’s efficiency in transmitting light, in terms of effective area in cm$^2$, 
\be
A_{\rm eff}
= A_{\rm coll}\times R_{\rm PM}\times R_{\rm SM}\times T_{\rm l} \times T_{\rm f}(\lambda)\times {\rm Q.E.}_{\rm det}(\lambda)\,,
\ee
where $A_{\rm coll}$ is the effective geometrical collecting area, $R_{\rm PM}$ and $R_{\rm SM}$ -- reflectivity of primary and secondary mirror in NUV, respectively, $T_{\rm f}(\lambda)$ and $T_{\rm l}$ -- filter and lens transmission in NUV, respectively, and Q.E.$(\lambda)$ is the quantum efficiency of the detector. A great advantage of AB magnitude system is that conversion to physical units is easy,
\be
F_{\nu}=3631\times 10^{-0.4m_{\rm AB}}\,\,\mbox(Ja)\,.
\ee
After considering the obstruction from the secondary baffle and secondary mirror holders, the geometrical collecting area is $A_{\rm coll}=37.724$ cm$^2$. Reflectivity of the Al-MgF$_2$-coated mirrors and transmission of the fused silica double-pass lens are assumed to be constant in the NUV (across the 200 -- 320 nm bandpass), 80\% and 90\%, respectively. Manufacture's filter transmission and detector quantum efficiency are presented in Fig.~\ref{fig:filter}, and the final effective area curve is presented in Fig.~\ref{fig:effective}. We also have estimated some of the photometric calibration constants: bandpass (the wavelength range over which the effective area is greater than 10\% of the peak), effective bandwidth $\D\l$, central (or mean) wavelength $\l_0$, pivot wavelength $\l_{\rm p}$, and effective wavelength $\lambda_{\rm eff}$ for three type of stars, Vega, A1V star and a white dwarf HZ 43 (presented in Table~\ref{table:constants}). All these values will be updated after the ground calibrations. 

\begin{table*}[ht]
\begin{center}
\caption{Preliminary values of the photometric calibration
constants in \AA. These values will be updated after the ground calibrations.} 
\label{table:constants}
\begin{tabular}{|c|c|c|c|c|c|c|}
\hline
Bandpass &$\D\l$ & Central $\l_0$ & Pivot $\l_{\rm p}$  &  \multicolumn{3}{|c|}{Effective $\lambda_{\rm eff}$}  \\
\hline
& & & & Vega & A1V star & HZ43  \\
\cline{5-7}
1965 -- 2900 &457.7   & 2441.1  & 2430.6 &  2426.4 &  2432.9 & 2549.3 \\
\hline
\end{tabular} 
\end{center}
\end{table*}

The real units of LUCI are electron counts that can be converted to flux units or magnitude by
\be
N(\text{electrons/sec}) = \int d\lambda F_{\lambda} A_{\rm eff}\,,
\ee
where $F_{\lambda}$ is the source flux in phot/sec/cm$^2$/\AA. Flux, corresponding to AB magnitude 12, is approximately $3.51\times 10^{-13}$ ergs/sec/cm$^2$/\AA. Total number of electrons generated in 10 sec exposure can be calculated by multiplying flux, effective area and exposure time, resulting in 472 electrons. The readout noise is 7 electrons/pix, and dark current is 2 electrons/pixel/sec at $25^{\circ}$C. Since the sky background in the NUV is very low, at 25 mag/arcsec \citep{uvbackgnd}, we consider it negligible for us. 

\begin{figure}[h]
\begin{center}
\includegraphics[scale=0.25]{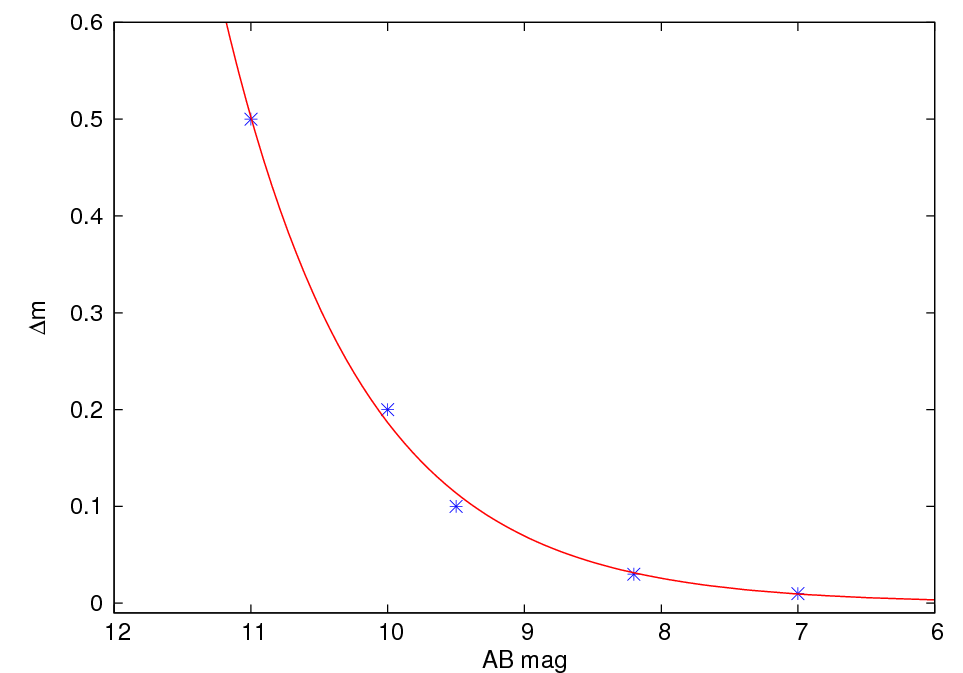} 
\end{center}
\caption{LUCI Photometric accuracy. Stars are calculated data points and the curve is an exponential fit.}
\label{fig:LUCI Photometric accuracy}
\end{figure}

The signal to noise ratio (SNR) can be calculated using the formula \citep{Herbert}, 
\be
SNR = \frac{S}{\sqrt{S+n\times\left(B+T+\sigma_{\rm R}^{2}\right)}}\,,
\ee
where $S$ is the total signal, $n$ is the number of pixels within the aperture, $B$ is the sky background, $T$ is the dark noise per pixel per second, and $\sigma_{\rm R}$ is the readout noise per pixel. Since 80\% of encircled energy falls within $4\times 4$ pixels, $n=16$ and $S = 80\%$ of the total energy. Therefore, from Eq.~(4) we find that in 10 sec exposure, $SNR = 4.34$. The full-well capacity of CCD is 14,000 electrons, and the CCD will saturate in the shortest exposure time (0.08 sec) corresponding to the signals from a star brighter than AB 2 mag and, hence, LUCI's bright limit is estimated to be AB 2 mag. Using these values, we can estimate LUCI's photometric accuracy in detecting the transients. For example, an AB 9.5 mag star, which is variable, a 10\% brightening or dimming of this star can  be detected by LUCI in 10 sec with SNR of 4. For different magnitude stars for 10 seconds exposure time, the photometric accuracy in detecting the brightness variation with SNR of 4 is plotted in Fig.~\ref{fig:LUCI Photometric accuracy}.

LUCI will be calibrated at the M.~G.~K. Menon Space  Science Centre at the CREST campus of the Indian Institute of Astrophysics, Bangalore, India. This facility was used for the integration, characterization and calibration of the UVIT instrument \citep{Kumar}. To avoid the reduction of the optics efficiency in the UV due to molecular and particulate contamination, we have selected low outgassing materials for LUCI and performed vacuum bake-out, ultrasonic cleaning and/or solvent cleaning at component level. Assembly of the  mirrors on the mount was performed in the Laboratory for Electro-Optics Systems (LEOS), ISRO, Bangalore; the assembled structure is stored in a class 1000 clean environment. In order to mitigate the impact of contamination, the telescope assembly and alignment will be performed in class 1000 clean rooms, with laminar flow tables providing class 100, or better, local environment. Procedures will be adapted from the UVIT calibration and, where possible, equipment from the UVIT program will be used. The complete instrument will be purged with ultra-clean nitrogen after the initial assembly.

The primary space photometric calibration of LUCI will be made with reference to a set of 
well-studied hot, young and/or massive stars, including bright $O$, $B$ and bright white 
dwarfs (WDs) as was done for the UV instruments onboard the HST. These stars are 
photometrically stable, their fluxes are well matched to the sensitivities of modern 
space instrumentation and their spectral energy distributions (SED) are well 
characterized. 

\section{Observation strategy}

LUCI will be mounted as a transit telescope, where it will look at  zenith to scan the sky in the NUV domain. The apparent motion of the celestial objects will allow the telescope scan a portion of the sky, and the observation will continue until the object moves out of the available sky region of the telescope (see details in paper~I). The total integrated exposure time on a single object is $2,300\times \sin{\psi}$ sec, where $\psi$ is the
altitude (this is the time it takes a point source to cross the centre of a $0.34^{\circ}$ FOV).

The default operation is with the zenithal orientation of LUCI (the ecliptic is always at $40^{\circ}$ from zenith), however, the inclination of up to $25^{\circ}$ from zenith can be imposed by the lander, depending on the final mission constraints. 

A majority of short-scale UV transients are stellar flares from late-type stars, for example, $K$ and $M$ dwarfs. $M$ dwarfs account for more than 75\% of the stellar population in the solar neighbourhood (up to 1 Kpc). The flares exceed the quiescent flux by 2 or 3 orders of magnitude with peak temperatures reaching $\sim 10^4$. For a thermal spectrum, a region of $2 - 6 \times 10^4$ K represents the NUV part of the spectrum. For such a flare on, for example, Proxima Cen at 1 pc away, this could correspond to a flux at Earth of $10^5 - 10^6$ UV phots/cm$^2$ for several hours. The M dwarf flare 100 pc away will produce UV flux at Earth of $10 - 10^2$ phots/cm$^2$, detectable by LUCI in 100 sec.

\begin{figure}[h!]
\begin{center}
\includegraphics[scale=0.45]{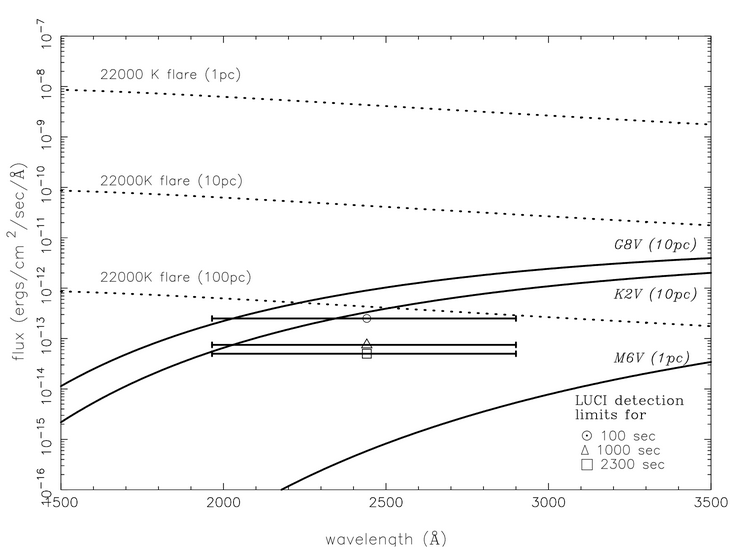}
\end{center}
\caption{LUCI detection limits for different integrated exposures.}
\label{fig:flare}
\end{figure}

In Fig.~\ref{fig:flare} we show the black body flux curves for a solar type star (G8V, $\tau$ Ceti taken as an example), a cooler star (K2V, $\epsilon$ Eri as an example) and an $M$ dwarf (M6V, Proxima Cen as an example) as seen from the same distances, along with UV sensitivity of LUCI for several exposure times with $SNR=5$. Fig.~\ref{fig:flare} also depicts the UV flares flux curves from a terrestrial-size hot spot with $T = 22000$ K as seen at different distances from Earth, resulting from possible collisions in extrasolar planetary systems, such as e.g. $\tau$ Ceti and $\epsilon$ Eri (see estimates of fluxes in Safonova, Sivaram \& Murthy, 2008). 

The required power for the detector will be provided by the lander, where solar panels are the primary source of electrical energy during lunar day surface operations, therefore LUCI will only operate in the daytime. The lander will provide a shared downlink data telemetry in addition to a limited command uplink facility; the total available data-link budget on the platform is 200 Kbps downlink and 100 bps uplink. Since LUCI is planned to be mostly contained within the lander, we will need only minimal thermal blankets, since the platform will take care of most of the insulation, and the situation is same for the radiation shielding. The telescope aperture will be well baffled to avoid reflections and scattering from the outer surfaces of the lander. In order to ensure the telescope's survival over the lunar night, the detector will be switched off, and the platform will have a minimal battery power to provide the heater.

\section{Conclusion and Future Development}

We have designed an all-spherical NUV telescope that can be flown on a range of platforms such as high-altitude balloons, nanosatellites, and space missions. An all-spherical design is a cost-effective alternative compared to other telescope configurations, reducing the manufacturing and alignment complexity. The structural analysis has shown that the instrument withstands all launch-load vibrations. We have met all major challenges in the development of the instrument, such as compactness, light weight and cost-effectiveness. This telescope will be flown aboard the Team Indus mission as part of Google Lunar X PRIZE competition at the end of 2017, and will perform the survey of the available sky from the lunar surface, primarily looking for bright UV transients.

As a qualification flight, we will fly this telescope on-board the high-altitude balloon in December  2016 to perform observations in the NUV domain from the floating altitude of about 40 km. Our team has developed the pointing and stabilization system for the balloon-born astronomical payloads \citep{Nirmal}, and the instrument will be mounted on this pointing platform for observations \citep{Sreejithballoon}. Another plan is to interface the in-house developed NUV spectrograph \citep{Sreejith} at the focal plane of the telescope, to carry out UV spectroscopy of bright sources. In addition, the telescope can be flown on any nanosatellite as a piggyback payload, where one of the main scientific objectives will be the study of the transient UV sky.

\section{Acknowledgments}
We would like to thank Dr.~R.~Sridharan,  Mr.~S.~Sriram, Mr.~S.~Nagabushana,Mr. Suresh Venkata, Mr.~P.~Umesh ~Kamath and Mr.~P.~K.~Mahesh of the Indian Institute of Astrophysics for their valuable suggestions. Part of this research has been supported by the Department of Science and Technology (Government of India) under Grant IR/S2/PU-006/2012.

\end{document}